\documentclass[12p]{iopart}

\usepackage[english]{babel}

\usepackage{braket,bbm,color,amssymb,amsbsy,graphicx,comment}
\usepackage[colorlinks,citecolor=blue,urlcolor=blue]{hyperref}
\usepackage{algorithm} 
\usepackage{algpseudocode}
\usepackage{array}
\usepackage{hhline}

\newcommand{\eqref}[1]{(\ref{#1})}

\newcommand{\Z}{\mathbbm{Z}}

\renewcommand{\i}{\mathrm{i}}

\newcommand{\abs}[1]{\left| #1 \right|}

\definecolor{mygreen}{rgb}{0,0.5,0}

\makeatother

\begin{document}

\title[Digital Quantum Simulation and Quantum Chaos in the Kicked Top]{Digital Quantum Simulation, Learning of the Floquet Hamiltonian, and Quantum Chaos of the Kicked Top}

\author{Tobias Olsacher$^{1,2,\footnotemark[1]}\footnotetext[1]{T.O. and L.P. contributed equally to this work.}$, Lorenzo Pastori$^{1,2,\footnotemark[1]}$, Christian Kokail$^{1,2}$, Lukas M. Sieberer$^3$, and Peter Zoller$^{1,2}$}

\address{$^1$ Center for Quantum Physics, University of Innsbruck, 6020 Innsbruck, Austria}

\address{$^2$ Institute for Quantum Optics and Quantum Information, Austrian Academy of Sciences, 6020 Innsbruck, Austria}

\address{$^3$ Institute for Theoretical Physics, University of Innsbruck, 6020 Innsbruck, Austria}

\date{\today}

\begin{abstract}
The kicked top is one of the paradigmatic models in the study of quantum chaos~[F.~Haake et al., \emph{Quantum Signatures of Chaos (Springer Series in
Synergetics vol 54)} (2018)]. Recently it has been shown that the onset of quantum chaos in the kicked top can be related to the proliferation of Trotter errors in digital quantum simulation (DQS) of collective spin systems. Specifically, the proliferation of Trotter errors becomes manifest in expectation values of few-body observables strongly deviating from the target dynamics above a critical Trotter step, where the spectral statistics of the Floquet operator of the kicked top can be predicted by random matrix theory. In this work, we study these phenomena in the framework of Hamiltonian learning (HL). We show how a recently developed Hamiltonian learning protocol 
can be employed to reconstruct the generator of the stroboscopic dynamics, i.e., the Floquet Hamiltonian, of the kicked top. We further show how the proliferation of Trotter errors is revealed by HL as the transition to a regime in which the dynamics cannot be approximately described by a low-order truncation of the Floquet-Magnus expansion. This opens up new experimental possibilities for the analysis of Trotter errors on the level of the generator of the implemented dynamics, that can be generalized to the DQS of quantum many-body systems in a scalable way. This paper is in memory of our colleague and friend Fritz Haake.
\end{abstract}

\maketitle


\section*{Preamble: Dedication to the Memory of Fritz Haake}

{\em This paper is dedicated to the memory of Fritz Haake. I remember vividly the Les Houches summer school 1995, where Fritz was teaching quantum chaos, and I taught a course on quantum optics in the early days of quantum information. As a younger generation quantum optics theorist, I had admired Fritz' contribution to theoretical quantum optics, and here we were, having breakfast together and talking physics, when Fritz pointed up to Mont Blanc and said: we should try to climb it. We spent the morning buying mountaineering equipment, and next morning we were on top of Mont Blanc, back to teaching early next day. This was Fritz, an enthusiastic and deep theoretical physicist, and an energetic sportsman, always ready for adventures and pushing limits, from science to sports. Our friendship continued, not only in off-piste skiing at the Obergurgl conferences and heli-skiing in Canada, but also in physics. It is now two years ago that we finished a paper together at the interface of quantum chaos and digital quantum simulation, and we will report on some newer developments below. Fritz will stay in our memory, not only as a gifted theoretical physicist and friend, but also as somebody who saw and lived science as an international effort, where  scientists are united by the common goal and endeavour to discover and understand, building bridges, and beyond any national boundaries and cultural identities. -- Peter Zoller }

\newpage

\section{Introduction} \label{sec:introduction}
\begin{figure} 
	\centering  
	\includegraphics[width=0.7\linewidth]{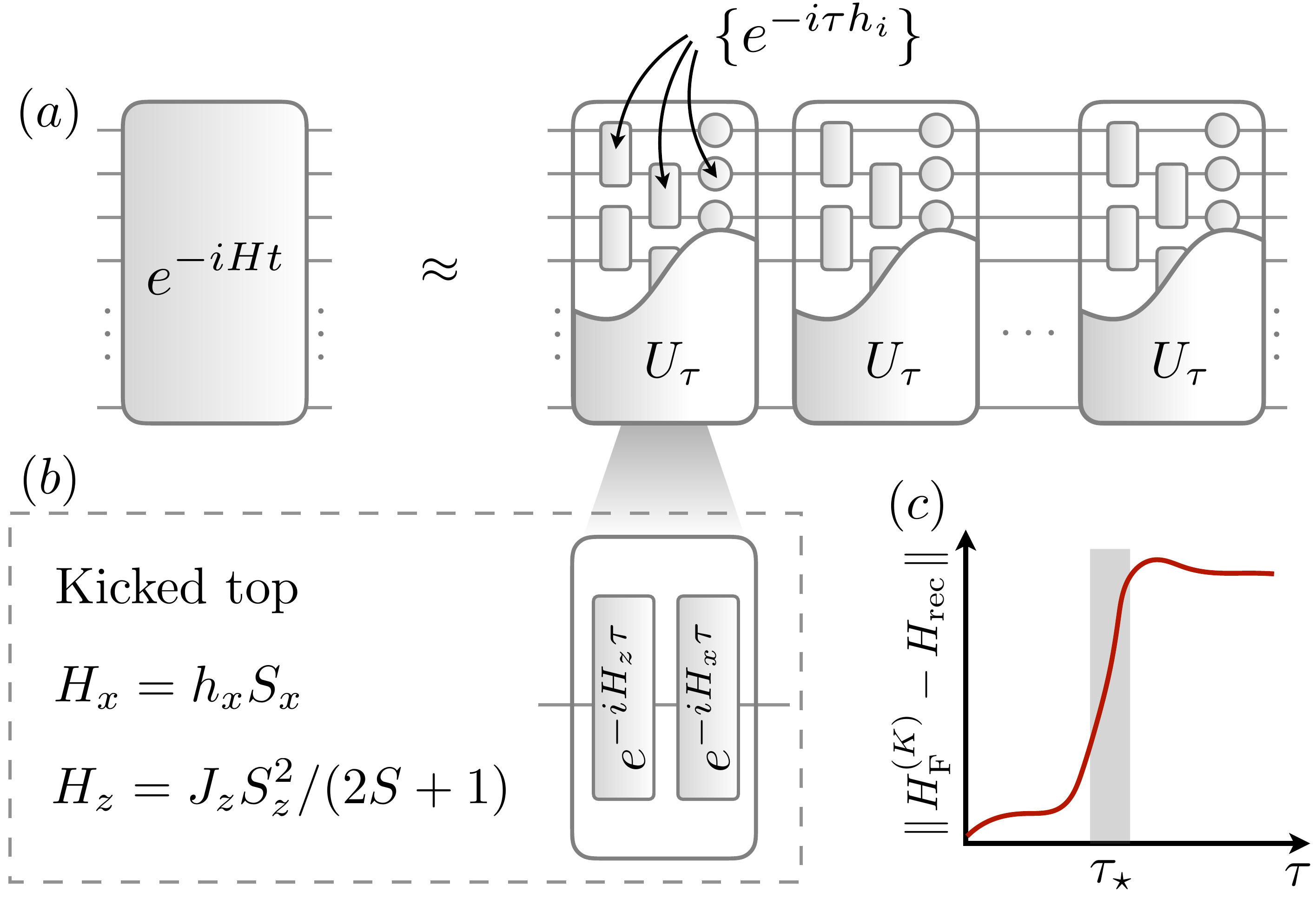}
	\caption{ \textbf{Digital quantum simulation (DQS) of quantum many-body systems, and quantum dynamics of the kicked top.} (a) DQS considers time evolution with a many-body Hamiltonian $H = \sum_i h_i$, which is approximated by a series of Trotter blocks $U_\tau$. These are constructed from elementary gate operations, e.g.~$\{ \mathrm{e}^{-\mathrm{i} \tau h_i}  \}$, so that  $U_{\tau} = \prod_i \mathrm{e}^{-\mathrm{i} \tau h_i}$. A single Trotter block can be expressed as $U_{\tau} = \mathrm{e}^{-\mathrm{i} H_{\rm F}(\tau) \tau}$, where $H_{\rm F}(\tau)$ denotes the Floquet Hamiltonian that, for sufficiently small $\tau$, can be written as a Floquet-Magnus expansion [see also Eq.~(\ref{eq:U-kicked-top}) and (\ref{eq:Fexp_Intro})]: $H_{\rm F}(\tau) = C_0 + \tau C_1 + \tau^2 C_2 + \cdots$. Here $C_0 = H$ is the desired target Hamiltonian, and the $C_{k>0}$ reflect Trotter errors consisting of higher-order commutators of the operators $h_i$ according to the Baker-Campbell-Hausdorff formula. (b) Trotter block in DQS for kicked top dynamics which consists of spin precession ($\mathrm{e}^{-\mathrm{i} H_x \tau}$) and non-linear kicks ($\mathrm{e}^{-\mathrm{i} H_z \tau}$) of a single spin. (c) Hamiltonian Learning (HL) provides an experimentally feasible protocol to reconstruct the Hamiltonian that governs the dynamics in a DQS experiment up to a given order $K$ in the Floquet-Magnus expansion, i.e, $H_{\rm F}^{(K)}=\sum_{k=0}^K\tau^k C_k$. The distance between the reconstructed Hamiltonian $H_{\mathrm{rec}}$ and $H_{\rm F}^{(K)}$ serves as a quantifier to study the Trotter threshold $\tau_{*}$ for the transition from regular to chaotic dynamics (shown here schematically).       }
  \label{fig:dqs}
\end{figure}

The kicked rotor and the kicked top as periodically driven quantum
systems represent paradigmatic models in studying quantum chaos (see Chapter 8 in \cite{Haake2018}), which have played a central role in the discussion of
phenomena like quantum localization and relation to random matrix theory (RMT). In recent collaborative work
with Fritz Haake~\cite{Sieberer2019} we have pointed out that the well-studied problem of
the transition from regular to chaotic dynamics, as observed in the kicked top as a function of the driving frequency, sheds new
light on and provides a physical interpretation of Trotter
errors in digital quantum simulation (DQS)---a highly relevant problem
in the focus of today's effort to `program' quantum many-body dynamics
on quantum computers. In DQS, the unitary evolution $U(t)=\mathrm{e}^{-\mathrm{i}Ht}$
generated by a Hamiltonian $H$ is simulated by decomposition
into a sequence of quantum gates \cite{Lloyd1996}. This can be achieved via a Suzuki-Trotter
decomposition~\cite{Trotter1959, Suzuki1976}, which approximately factorizes the time evolution operator
in Trotter time steps of size $\tau$ [see Fig.~\ref{fig:dqs}(a)]. 
While the `correct' evolution
operator $U(t)$ will emerge in the limit of Trotter stepsize $\tau\rightarrow0$,
in practice the finite fidelity of quantum gates makes it desireable
to take as large Trotter steps as possible~\cite{Knee2015}.
The observation of Refs.~\cite{Sieberer2019,Heyl2019,Kargi2021} was that the error associated with a finite Trotter
step size shows a sharp threshold behavior: while for a small time
step $\tau<\tau_{*}$ Trotterized time evolution provides a faithful
representation of the desired dynamics, in the regime $\tau>\tau_{*}$
Trotter errors proliferate. This behavior is in correspondence to
a transition from regular motion to quantum chaos in Floquet systems when the driving
frequency is decreased \cite{DAlessio2014,Regnault2016}. While the ultimate goal of DQS is to simulate complex quantum many-body systems with finite range
interactions (see recent advances in Refs.~\cite{Peng2005,Lanyon2011,Barreiro2011,Weimer2011,Wecker2014,Poulin2014,Barends2015, Salathe2015, Barends2016, Langford2016,OMalley2016, Martinez2016,Seetharam2021}), the features of a Trotter threshold
are already visible in simple models. This leads us back to the kicked top, which---while being intrinsically a single particle problem (for a `large' spin $S$)---can also be interpreted
as Trotterized time evolution of a many-body spin-model with infinite
range interactions. Thus, the kicked top can serve as a testing ground, both theoretically
and experimentally, for the phenomenology of the Trotter threshold. 

In this work, we discuss the Trotter threshold from the perspective of `Hamiltonian learning' (HL), and we choose the kicked top as a simple model system displaying pertinent features. The discussion is based on the HL framework for the characterization of Trotterized DQS of many-body systems developed by us in Ref.~\cite{Pastori2022}~\footnote{The protocol developed in \cite{Pastori2022} further extends to the learning of Liouvillians, for the characterization of dissipative dynamics.}, which we adapt and extend here to study the Trotter threshold and the transition to quantum chaos in the kicked top~\cite{Sieberer2019}. HL is presently being developed as a new tool in quantum
information theory in the context of quantum many-body systems and quantum simulation~\cite{Garrison2018,Chertkov2018,Qi2019,Hou2020,Bairey2019,Bairey2020,Li2020,Evans2019,Zubida2021,Anshu2021,Eisert2020,Carrasco2021,Bienias2021}. In the
present context, we can phrase HL as follows: we 
consider quench dynamics of a many-body spin system, $\ket{\psi(t)}=\mathrm{e}^{-\mathrm{i}Ht}\ket{\psi(0)}$,
where the system is initially prepared in the state $\ket{\psi(0)}$
and evolves under a (time-independent) Hamiltonian $H$ to a final
state $\ket{\psi(t)}$ at time $t$. The goal is to learn the operator
content of $H$, i.e., a tomographic reconstruction of $H$ from measurements
on $\ket{\psi(t)}$. The key to an {\em efficient} learning of $H$ from
experimental observations is that a \emph{physical} many-body Hamiltonian
consists of a small (polynomial) number of terms, i.e., the operator content of $H$ will be limited to one-body, quasi-local two-body terms
{\em etc.}, while the many-body wave function lives in a Hilbert space of dimension scaling exponentially with the number of constituents. HL thus becomes efficient
by having to learn only a sufficiently small number of coupling coefficients in the Hamiltonians,
while testing for presence of additional terms, and thus verifying the learned Hamiltonian
structure with more data. 
In the following we apply these ideas to the kicked top, viewed as DQS of a collective spin system.




The kicked top combines precession of the spin $S$ of the top around the $x$-axis with $\tau$-periodic non-linear ``kicks'' around the $z$-axis, according to the time-dependent Hamiltonian
 \begin{equation}
   \label{eq:H-kicked-top}
   H_\mathrm{KT}(t) = H_x + \tau H_z \sum_{n \in \Z} \delta(t - n \tau),
 \end{equation}
where $H_x = h_x S_x$ and $H_z = J_z S_z^2/(2 S + 1)$ with quantum angular momentum operators $S_{\mu}$ with $\mu = x, y, z$. The evolution operator generated by $H_\mathrm{KT}(t)$ over a single period of duration $\tau$ can be equivalently described in terms of a Floquet operator
 \begin{equation}
   \label{eq:U-kicked-top}
   U_{\tau} = \e^{-\i H_z \tau} \e^{-\i H_x \tau}\equiv \e^{-\i H_{\rm F} (\tau)\tau},
 \end{equation}
as illustrated in Fig~\ref{fig:dqs} (b). The dynamics of the kicked top is quantum chaotic iff the spectral statistics of $U_\tau$ can be described by one of Dysons's ensembles of random matrices~\cite{Haake2018}. Indeed, while random matrix theory (RMT) was initially applied in physics by E.~Wigner to understand the distribution of nuclear spectra (see \cite{Weidenmueller2009} for a review on RMT in nuclear physics), it forms now the basis for the study of quantum chaos. It is indeed a defining feature of quantum chaotic
systems that their spectral statistics are universal and obey predictions from random-matrix theory (RMT)~\cite{Haake2018,Guhr1998}. 

 Alternatively Eq.~\eqref{eq:U-kicked-top} constitutes the elementary gate sequence of a DQS that aims at approximating the Hamiltonian $H=H_z + H_x$ according to $\mathrm{e}^{-\mathrm{i}Ht} \approx U_{\tau = t/n}^n$, where $t$ denotes the total simulation time which is split into $n$ steps of duration $\tau=t/n$. We emphasize that the accuracy of this approximation does not only depend on the Trotter step $\tau$, but also on the Trotter sequence that can be chosen to compensate Trotter errors up to a given order $\mathcal{O}(\tau^k)$~\cite{Wiebe2010,Chen2022}.
Eq.~\eqref{eq:U-kicked-top} also defines the Floquet Hamiltonian $H_{\mathrm{F}} (\tau)$. For sufficiently small $\tau$, the Floquet Hamiltonian can be written as a Floquet-Magnus expansion, i.e., employing Baker–Campbell–Hausdorff formulas,
\begin{equation} \label{eq:Fexp_Intro}
 H_{\mathrm{F}} (\tau) = H_x + H_z + \mathrm{i}\,\frac{\tau}{2} [H_x,H_z] + \ldots
\end{equation}
which is a series expansion in the Trotter stepsize $\tau$, written here up to first order, with the higher order terms taking the form of nested commutators of $H_x$ and $H_z$ (see Sec.~\ref{sec:Review} below). The question of convergence of this series is intrinsically connected with the transition from regular to chaotic dynamics at a specific $\tau_{*}$. From a DQS point of view, in the limit $\tau\rightarrow 0$, the Floquet Hamiltonian reduces to $H=H_z+H_x$, as the desired Hamiltonian `to be simulated' on a quantum device. Higher order terms in $\tau$ represent Trotter errors. 

While in traditional discussions of DQS, Trotter errors are quantified in terms of Trotter bounds, $|| U_\tau - \mathrm{e}^{-\mathrm{i}H\tau}|| \equiv {\cal O} (\tau^k)$ \cite{Suzuki1985,Childs2019,Childs2020}, HL goes beyond in quantifying these errors by learning terms order by order of the Floquet-Magnus expansion in an experimentally feasible protocol \cite{Pastori2022}. Central to  our work below is the toolset built into HL which quantifies errors of an {\em Ansatz Hamiltonian}, e.g., as a truncated FM series Eq.~\eqref{eq:Fexp_Intro}, to represent the experimental $ H_{\mathrm{F}} (\tau)$. This will be our key quantifier in studying the Trotter threshold $\tau_{*}$ [see Fig.~\ref{fig:dqs} (c)]. Thus our work goes significantly beyond Refs.~\cite{Sieberer2019,Heyl2019,Kargi2021}, where the Trotter threshold was studied only for low-order observables. While our discussion below will focus on the kicked top as a simple model system, we emphasize that the main results and conclusions carry over to characterizing Trotter errors in DQS of quantum many-body systems from many-body models in condensed matter physics, to quantum chemistry or high energy physics.

This paper is organized as follows. In Sec.~\ref{sec:Review} we briefly summarize our previous work~\cite{Sieberer2019} on the Trotter threshold of the kicked top. Sec.~\ref{sec:HLinKickedTop} provides the main results of the present work. We will start with a description of HL protocols to learn order by order, up to given truncation cutoff $K$, the Floquet Hamiltonian of the kicked top, followed by a discussion of numerical results illustrating the technique. We conclude with Sec.~\ref{sec:Outlook}.

\section{The Trotter Threshold Revisited} \label{sec:Review}

In preparation for the discussion in Sec.~\ref{sec:HLinKickedTop} on HL applied
to the Floquet Hamiltonian for the kicked top, we start by reviewing previous
work, and, in particular, our collaborative work with Fritz
Haake~\cite{Sieberer2019}.

\subsection{Quantum Many-Body Models and the Kicked Top} \label{subsec:TrotterThreshQMB}

The motivation for the present study is the quantitative characterization of Trotter errors, and the Trotter threshold in particular, in quantum many-body systems. Therefore, we find it useful to recall some of the basic features of DQS of quantum many-body systems, which we illustrate here for 1D spin models, in relation to the kicked top as the model system studied below. 

Typical model
systems of interest are one-dimensional chains of $N$ spin-$1/2$ such
as the long-range Ising model with Hamiltonian
\begin{equation}
  \label{eq:H-Ising}
  H = H_x + H_z, \quad H_x = h_x \sum_{i = 1}^N \sigma_i^x, \quad H_z = J_z
  \sum_{i < j = 1}^N \frac{\sigma_i^z \sigma_j^z}{\abs{i - j}^{\alpha}},
\end{equation}
where $\sigma_i^{\mu}$ with $\mu = x, y, z$ are Pauli operators for spins on
lattice sites $i = 1, \dots,
N$. Power-law interactions with $0 \leq \alpha \leq 3$ are routinely implemented with trapped-ion quantum simulators~\cite{Monroe2021}. The kicked top emerges in DQS in the limit $\alpha \rightarrow 0$.

Before proceeding to a study of the kicked top, we recall some of the basic features of Trotter errors and the Trotter threshold, which have emerged in our previous work. In DQS, time evolution generated by $H$ is represented by a sequence of
elementary quantum gates. This is achieved through the approximate factorization
of the time evolution operator within each Trotter step, $\e^{- \i H \tau} \approx \e^{-\i H_z \tau} \e^{- \i H_x \tau}$: Individual terms in the sums in $H_x$ and
$H_z$, respectively, commute with each other, so that $\e^{- \i H_x \tau}$ and
$\e^{- \i H_z \tau}$ can directly be decomposed into single-spin and two-spin
gates. However, due to the non-commutativity of the components of the target
Hamiltonian, $[H_x, H_z] \neq 0$, the factorization of the time evolution
operator within a single Trotter step is exact only in the limit $\tau \to 0$,
and any finite value $\tau > 0$ leads to the occurrence of Trotter errors. On
the level of the generator of the dynamics, the FM expansion
Eq.~\eqref{eq:Fexp_Intro} suggests that Trotter errors are perturbatively small
in $\tau$. However, a rigorous sufficient condition for the convergence of the
FM expansion~\cite{Casas2007, Moan2008, Blanes2009} indicates that the radius of
convergence scales with system size as $\sim 1/N$, which would imply that the FM expansion is not
applicable in the thermodynamic limit $N \to \infty$, and puts into question whether Trotter errors can be controlled in DQS of quantum many-body systems. Addressing this question requires a suitable measure of Trotter errors. In DQS, quantities of physical interest are typically expectation values of few-body
observables, i.e., (sums of) products of spin-$1/2$ operators acting on only a
few different spins, such as $\sigma_i^{\mu}$, $\sigma_i^{\mu} \sigma_j^{\nu}$,
$\sigma_i^{\mu} \sigma_j^{\nu} \sigma_k^{\rho}$, etc. Therefore, it is natural to quantify Trotter errors in terms of deviations of expectation values of few-body observables from their target values that are obtained by time evolution generated by the target Hamiltonian $H$. For Ising spin chains, in the limit
$\alpha \to \infty$ of short range interactions and for $0 \leq \alpha \leq 3$,
such quantitative studies of Trotter errors of few-body observables were carried
out in Refs.~\cite{Heyl2019} and~\cite{Sieberer2019}, respectively. As we will
illustrate with a concrete example below, these studies found sharp threshold
behavior, where Trotter errors remain controlled for small Trotter steps and
proliferate for $\tau$ larger than a threshold value $\tau_{*}$.

The Trotter threshold was observed consistently over the entire range of values
of power-law interaction exponents $\alpha$ considered. In the limit
$\alpha \to 0$, DQS of the spin model in Eq.~\eqref{eq:H-Ising} is directly
related to the dynamics of a kicked top: For $\alpha = 0$, the components of the
Ising Hamiltonian in Eq.~\eqref{eq:H-Ising} can be cast, in terms of collective
spin operators $S_{\mu} = \sum_{i = 1}^N \sigma_i^{\mu}/2$, as $H_x \sim S_x$ and
$H_z \sim S_z^2$---just as in the Hamiltonian of the kicked top in
Eq.~\eqref{eq:H-kicked-top}. Then, the collective spin
$\boldsymbol{S}^2 = S_x^2 + S_y^2 + S_z^2$ becomes a constant of motion. Consequently, the
many-body Hilbert space with dimension $2^N$ is decomposed into decoupled
subspaces of fixed total spin $S$, and within each subspace, the Trotterization
of Eq.~\eqref{eq:H-Ising} reproduces the dynamics of a kicked top of size
$S$. In this sense, the kicked top becomes a single-particle (i.e., a single collective spin) toy model for the many-body
Trotter threshold, where the notion of few-body observables introduced above to
quantify Trotter errors translates to low-order products of spin operators
$S_{\mu}$. In the following, we give a detailed account of the Trotter threshold
in the context of the kicked top.

\subsection{The Kicked Top and the  Trotter Threshold}

\subsubsection{Model System and Magnus Expansion}

Before we enter a detailed discussion of the Trotter threshold, we introduce the
following extension of the Floquet operator given in
Eq.~\eqref{eq:U-kicked-top}:
\begin{equation}
  \label{eq:U-CUE}
  U_{\tau} = \mathrm{e}^{-\mathrm{i} H_z \tau} \mathrm{e}^{-\mathrm{i} H_y \tau}
  \mathrm{e}^{-\mathrm{i} H_x \tau} = \e^{- \i H_{\mathrm{F}}(\tau) \tau}.
\end{equation}
This extended Floquet operator, on which our discussion of the Trotter threshold
will focus, corresponds to DQS of a target Hamiltonian $H = H_x + H_y + H_z$,
where
\begin{equation}
    H_{\mu}=\frac{J_{\mu}S^2_{\mu}}{2S+1}+h_{\mu}S_{\mu} \,\,.
\end{equation}
The rationale behind this choice of model, as opposed to the model for the kicked top in the introduction,  is explained in \ref{app:vari-kick-top}: the key difference between the Floquet operators in Eqs.~\eqref{eq:U-kicked-top}
 and~\eqref{eq:U-CUE} are the absence of time-reversal and geometrical symmetries as well as resonant driving points in the latter
case~\cite{Haake2018}. We choose $J_z$ as the unit of energy, and, for
 concreteness, we fix $J_y = 0$, $J_x = 0.4 J_z$, $h_z = h_y = 0.1 J_z$, and $h_x = 0.11 J_z$. 
Further, we note that while the Floquet-Magnus (FM) expansion always takes the form of a power series in the Trotter step size $\tau$,
\begin{equation} \label{eq:Fexp}
	H_\mathrm{F}(\tau) = \sum_{k=0}^\infty \tau^k\,C_k \,\,,
\end{equation}
for the specific case of a three-step Floquet drive in Eq.~\eqref{eq:U-CUE}, the first few terms are given by
\begin{eqnarray} \label{eq:FexpTerms}
    C_0 & = \sum_\alpha H_\alpha\,\,, \label{eq:FexpTerms1}\\ 
    C_1 & =  \frac{\mathrm{i}}{2} \sum_{\alpha < \beta} [H_\alpha,H_\beta]\,\,, \label{eq:FexpTerms2}\\ 
    C_2 & = -\sum_{\alpha \neq \beta}\frac{[H_\alpha, [H_\alpha, H_\beta]]}{12} - \frac{[H_x,[H_y,H_z]]}{6} - \frac{[H_z,[H_y,H_x]]}{6}\,\,, \label{eq:FexpTerms3}
\end{eqnarray}
where $\alpha, \beta \in \{x, y, z\}$.

\subsubsection{Quantifying Trotter Errors }

Among different choices of few-body observables to quantify Trotter errors, a
special role is played by the target Hamiltonian itself: for $\tau \to 0$, the
Floquet operator in Eq.~\eqref{eq:U-CUE} reduces to time evolution generated by
the time-independent target Hamiltonian $H$, and the energy as measured by the
expectation value of $H$ becomes a constant of motion. To quantify the degree
to which this conservation law is obeyed in DQS, we define the simulation
accuracy~\cite{Bukov2016, Heyl2019}:
\begin{equation}
  \label{eq:QE}
  Q_E(n \tau) = \frac{E_{\tau}(n \tau) - E_0}{E_{T = \infty} - E_0}.
\end{equation}
Energy is conserved if $Q_E(n \tau) = 0$, and the energy at time $t = n \tau$,
given by
$E_{\tau}(n \tau) = \braket{\psi(0) | U_{\tau}^{\dagger n} H U_{\tau}^n |
  \psi(0)}$,
equals the energy of the initial state, $E_0 = \braket{\psi(0) | H | \psi(0)}$.
In contrast, the value $Q_E(n \tau) = 1$ indicates that the system absorbs
energy from the time-periodic Floquet drive and heats up to infinite temperature
$T = \infty$ such that $E_{\tau}(n \tau) = E_{T = \infty} = \tr(H)/\mathcal{D}$
where $\mathcal{D} = 2 S + 1$ is the Hilbert space dimension.

\subsubsection{Trotter Threshold}
\label{subsubsec:TrottThresh}
\begin{figure} 
	\centering  
	\includegraphics[width=\linewidth]{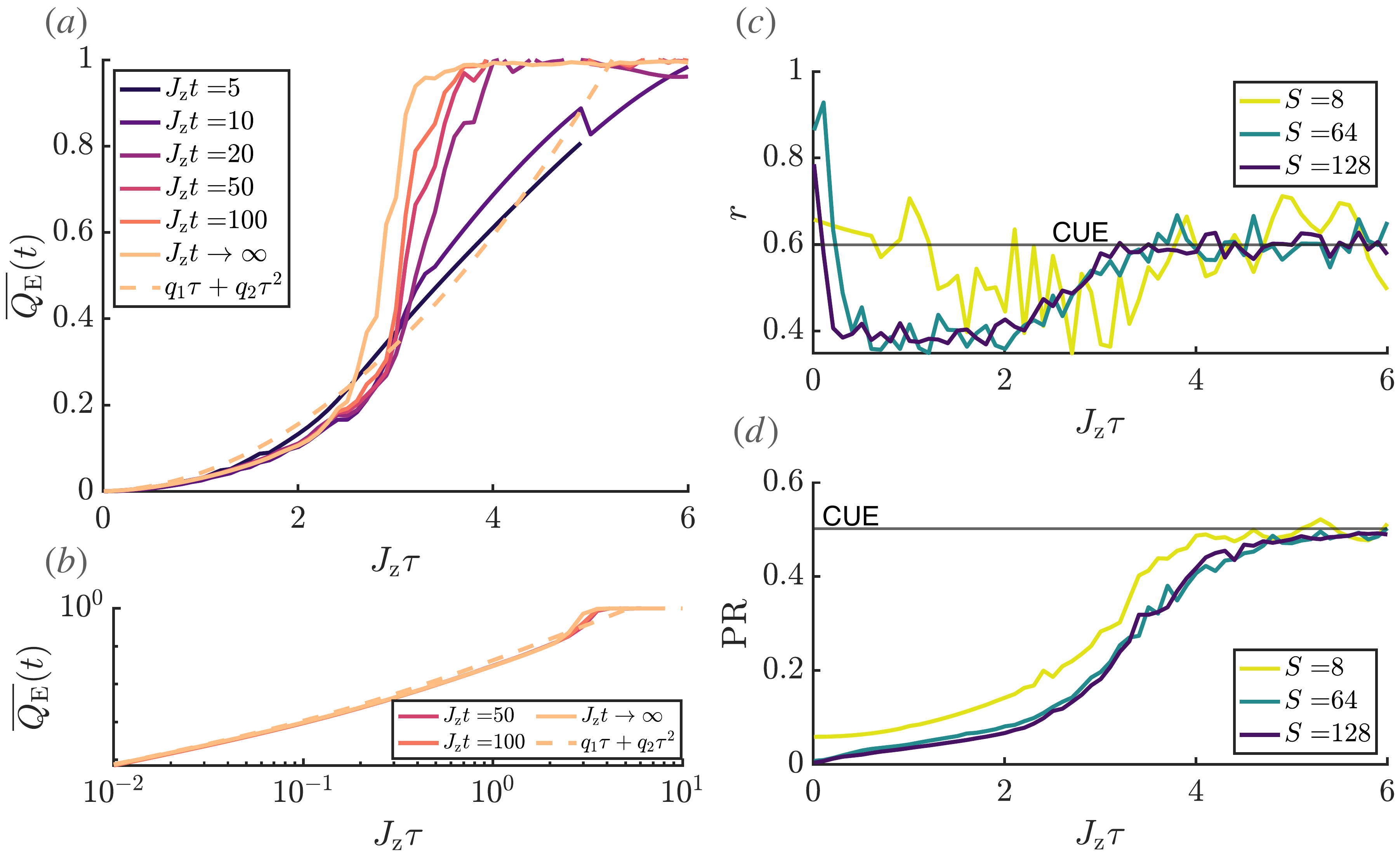}
	\caption{{\bf{Trotter threshold and transition to quantum
        chaos.}} (a)\&(b) The temporal average of the simulation accuracy
     Eq.~\eqref{eq:QE} exhibits threshold behavior for $J_z t \gtrsim 20$ ($S=128$ for all lines). For $t \to \infty$ and below the threshold, the numerical data is well-reproduced by the expansion to second order in $\tau$ given in Eq.~\eqref{eq:QE-F-expansion}, which is obtained from time-dependent perturbation theory based on a low-order truncation of the FM expansion~\cite{Sieberer2019}. The Trotter threshold coincides with the onset of quantum chaos: for $\tau \gtrsim \tau_{*}$, (c) the adjacent phase spacing ratio Eq.~\eqref{equ:AVGphaserepulsion} and (d) the participation ratio Eq.~\eqref{equ:PR} converge, for $S \to \infty$, to the CUE values $r_{\mathrm{CUE}} \approx 0.6$ and $\mathrm{PR}_{\mathrm{CUE}} = 1/2$~\cite{Ullah1963}, respectively. Below the Trotter threshold the adjacent phase spacing ratio in (c) is governed by Poisson statistics leading to $r_\mathrm{POI}\approx 0.39$. 
     }
  \label{fig:threshold-chaos}
\end{figure}

The temporal average
$\overline{Q_E}(t) = \frac{1}{n_t} \sum_{n = 1}^{n_t} Q_E(n \tau)$, where
$n_t = \left\lfloor t/\tau \right\rfloor$ is the number of Trotter steps
corresponding to the simulation time $t$, is shown in
Fig.~\ref{fig:threshold-chaos}(a). Here, the initial state is chosen as a spin coherent state,
$\ket{\theta, \phi} = \e^{\i \theta \left( S_x \sin(\phi) - S_y \cos(\phi)
  \right)} \ket{S, S_z = S}$,
with $\theta = 0.1$ and $\phi = 0.2$. At times $J_z t \gtrsim 20$, Trotter errors, 
quantified by the time-averaged simulation accuracy $\overline{Q_E}(t)$, exhibit
threshold behavior: while $\overline{Q_E}(t)$ increases smoothly for small
values of $\tau$, a sudden jump to the saturation value
$\overline{Q_E}(t) \approx 1$ occurs at $J_z \tau_{*} \approx 3.5$. The Trotter threshold 
persists for $t \to \infty$, and becomes sharper with increasing spin size
$S$~\cite{Sieberer2019}.
An analogous Trotter threshold can be observed also for other few-body
observables~\cite{Sieberer2019}, and, as pointed out above, applies also to
one-dimensional spin chains with algebraic~\cite{Sieberer2019} and
nearest-neighbor~\cite{Heyl2019} interactions, as well as bosonic models with
infinite Hilbert spaces~\cite{Kargi2021}. It should be noted that generic and
sufficiently short-range interacting many-body systems are expected to heat up
indefinitely when subjected to Floquet driving~\cite{DAlessio2014b,
  Lazarides2014, Luitz2017}, so that the Trotter threshold is washed out with
$\overline{Q_E}(t) \to 1$ for any $\tau > 0$ at late times. However, the heating
rate is exponentially small in the driving frequency~\cite{Abanin2015, Mori2016,
  Kuwahara2016, Abanin2017, Machado2019, Howell2018, Rakcheev2020}. Thus, for
given finite simulation run-time, heating can be suppressed efficiently.

Below the Trotter threshold, Trotter errors remain controlled, i.e., perturbatively
small in $\tau$. In fact, Trotter errors are well-described by a low-order
truncation of the FM expansion: treating corrections to the target Hamiltonian
in Eq.~\eqref{eq:Fexp} in time-dependent perturbation theory~\cite{Heyl2019,
  Sieberer2019}, the long-time average of the simulation accuracy,
$\overline{Q_E} = \lim_{t \to \infty} \overline{Q_E}(t)$, can be written as a
power series in $\tau$:
\begin{equation}
\label{eq:QE-F-expansion}
  \overline{Q_E} = q_1 \tau + q_2 \tau^2 + \mathcal{O}(\tau^3),
\end{equation}
with, for the parameters chosen for Fig.~\ref{fig:threshold-chaos}(a), 
$q_1 \approx 0.007$ and $q_2 \approx 0.035$. The good agreement between this series
expansion and the numerical data for $\tau \lesssim \tau_{*}$ shows that a
low-order truncation of the FM expansion provides a quantitatively accurate
description of the dynamics of few-body observables well beyond rigorous bounds
on the radius of convergence of the FM expansion.

\subsubsection{Breakdown of the Floquet-Magnus Expansion and Onset of Quantum Chaos} \label{subsubsec:quantumchaoskickedtop}

The breakdown of the FM expansion, signalled by the proliferation of Trotter errors for
$\tau > \tau_{*}$, marks the onset of quantum chaos. As a first indication for
the connection between the divergence of the FM expansion and quantum chaos, we
note that the saturation value $\overline{Q_E} = 1$ at $\tau \gtrsim \tau_{*}$
is consistently reproduced by replacing $U_{\tau}$ in each Trotter step by a random unitary
matrix. Then, the temporal average is manifestly equivalent to a Hilbert-space average,
which by definition yields the infinite-temperature value $E_{T = \infty}$.

More systematically, quantum chaos of the Floquet operator $U_{\tau}$ manifests
in statistics of the eigenphases $\theta_n$ and in localization properties of
the eigenvectors $\ket{\phi_n}$, which obey the eigenvalue equation
$U_{\tau} \! \ket{\phi_n} = \e^{\i \theta_n} \! \ket{\phi_n}$. In particular, as a \emph{defining} signature of quantum chaos, the distribution of spacings of eigenphases $\theta_n$ is described by RMT, and the eigenvectors $\ket{\phi_n}$ are delocalized in a basis of eigenvectors of an operator that generates integrable dynamics, such as the target Hamiltonian. In the following, we study the onset of chaos in the kicked top via comparison of the eigenphases and eigenstates of the Floquet operator to their respective RMT predictions. We refer the reader to Ref.~\cite{Haake2018} for a comprehensive introduction to the signatures of quantum chaos.

Eigenphase statistics are conveniently characterized by the average adjacent phase spacing
ratio $r$~\cite{Oganesyan2007},
\begin{equation}
  \label{equ:AVGphaserepulsion}
  r =\frac{1}{\mathcal{D}} \sum_{n=1}^\mathcal{D} r_n, \qquad r_n = \frac{\min(\delta_n,
    \delta_{n+1})}{\max(\delta_n, \delta_{n+1})},
\end{equation}
where $\delta_n = \theta_{n+1} - \theta_n$. This quantity constitutes a measure of the degree of repulsion between the eigenphases of $U_\tau$, and takes characteristic values for unitaries that are drawn from an  ensemble of RMT. In particular, for the circular
unitary ensemble (CUE) of random unitary matrices, the average adjacent phase spacing
ratio is given by $r_{\mathrm{CUE}} = 0.5996(1)$~\cite{Atas2013}. Indeed, this
value is reproduced for $\tau \gtrsim \tau_{*}$ as shown in
Fig.~\ref{fig:threshold-chaos}(b), giving a clear indication for quantum-chaotic dynamics beyond the Trotter threshold. At very small values $J_z \tau \lesssim 1/S$, the
phase spacing ratio is determined by the target Hamiltonian. 
The phase spacing
ratio drops to $r_\mathrm{POI} \approx 0.39$ for $1/S \lesssim J_z \tau \lesssim J_z \tau_{*}$, which
signals the absence of level repulsion in crossings of eigenphases that wind
repeatedly around the unit circle, leading to Poisson statistics for the adjacent phase spacings~\cite{Sieberer2019}. 
In this regime,
Trotterization leads to only weak mixing between the eigenvectors $\ket{\psi_n}$
of the target Hamiltonian, and the Floquet states $\ket{\phi_n}$ are in
one-to-one correspondence with the states $\ket{\psi_n}$. This localization of
Floquet states in the eigenbasis of the target Hamiltonian is quantified by the
participation ratio (PR),
\begin{equation}
  \label{equ:PR}
  \mathrm{PR} = \left( \sum_{n,m = 1}^{\mathcal{D}} \vert \langle \psi_n \vert
    \phi_m \rangle \vert^4 \right)^{-1}.
\end{equation}
The participation ratio is shown in Fig.~\ref{fig:threshold-chaos}(c). The small
values of PR at $\tau \lesssim \tau_{*}$ indicate a high degree of similarity
between the eigenbases of $H$ and $U_{\tau}$. For $\tau \gtrsim \tau_{*}$, the PR
saturates to a large value that indicates equal absolute overlaps between all
eigenstates, as expected for the eigenvectors $\ket{\phi_n}$ of a random matrix
$U_{\tau}$. In this regime, the precise numerical value of the PR is determined by the corresponding ensemble of RMT. For the CUE, and in the limit $\mathcal{D} \to \infty$, it is given by $\mathrm{PR}_{\mathrm{CUE}} = 1/2$~\cite{Ullah1963}.

\section{Learning the Floquet Hamiltonian $H_{\rm F}(\tau)$ of the Kicked Top} \label{sec:HLinKickedTop}

Previous work, as summarized in Sec.~\ref{sec:Review}, has focused on the Trotter
threshold  by monitoring
the simulation accuracy $Q_\mathrm{E}$, participation ratio, and the level spacing ratio of the Floquet operator $U_{\tau}$. Instead, the present section
will study the phenomenology of the Trotter threshold, and the transition to quantum chaos, from the view
point of {\em learning} the Floquet Hamiltonian $H_{\mathrm{F}}(\tau)$
in the form of the Floquet-Magnus expansion Eq.~\eqref{eq:Fexp}, as a function of $\tau$. 

Below we first describe
the technique of HL (Sec.~\ref{subsec:HL}), which we then adapt to `learn'
the Floquet-Magnus expansion of the kicked top. Central
to our discussion is the ability of the HL protocol to provide a quantitative
assessment of errors in HL, in particular in learning $H_{\mathrm{F}}(\tau)$
with $\tau$ approaching the Trotter threshold $\tau_{*}$.  Corresponding results for the kicked top will be presented in Sec.~\ref{subsec:resultsHL}.

\subsection{Hamiltonian Learning} \label{subsec:HL}

\subsubsection{Hamiltonian Learning Protocols for Quantum Many-Body Systems}

Recently, HL has been developed as a technique to efficiently recover
an unknown Hamiltonian of an isolated quantum many-body system via
measurements on quantum states prepared in the laboratory. Motivation
for HL is provided by the ongoing development of controlled quantum
many-body systems, e.g., as analog quantum simulators, where HL serves
to characterize and thus verify the functioning of quantum devices.
Various methods for recovering the Hamiltonian have been described,
from learning the Hamiltonian from a single eigenstate~\cite{Garrison2018,Chertkov2018,Qi2019,Hou2020}, or stationary
states~\cite{Bairey2019,Evans2019}, or from observation of short time dynamics~\cite{Li2020,Zubida2021}. 

To be specific, we give a brief description of HL in quench dynamics.  We focus
on HL as proposed in Ref.~\cite{Li2020} as our discussion below, on learning
 the Floquet Hamiltonian $H_{\mathrm{F}}(\tau)$, directly builds on
this method. Reference \cite{Li2020} considers quench dynamics of an isolated
many-body system, $\ket{\psi(t)}=\e^{-\i Ht}\ket{\psi(0)}$, where an initial
state $\ket{\psi(0)}$ evolves in time $t$ according to a \emph{time-independent} Hamiltonian $H$ to a
final state $\ket{\psi(t)}$. The task is to recover the operator structure of
$H$ by measuring observables at various times $t$. The key to an \emph{efficient}
reconstruction is the locality of the physically implemented many-body
Hamiltonian. This means that $H$ can typically be expanded as a sum of few-body operators, and can thus be specified by a number of coupling parameters that scales at most polynomially in system size $N$. In the specific
example of the long-range interacting spin-$1/2$ model discussed in Sec.~\ref{sec:Review} [see Eq.~\eqref{eq:H-Ising}], of which the kicked top can be seen as the limit $\alpha\to 0$, the Hamiltonian is expressed as a sum of Pauli operators $\sigma_i^z\sigma_j^z$ with coefficients $J^z_{i,j}\propto\frac{J_z}{|i-j|^{\alpha}}$, and depends thus on $\mathcal{O}(N^2)$ coupling coefficients, far less than the $\mathcal{O}(4^N)$ parameters needed to express a generic operator.

We thus seek to reconstruct the physically implemented Hamiltonian from an \emph{Ansatz}
\begin{equation}
  \label{eq:H-ansatz}
  H(\boldsymbol{c}) = \sum_{h_j \in \mathcal{A}} c_j h_j \,\,,
\end{equation}
specified by a chosen set $\mathcal{A}=\{h_j\}_{j=1}^{N_{\mathcal{A}}}$ of $N_{\mathcal{A}}=\mathrm{poly}(N)$ few-body Ansatz operators $h_j$, and depending on coefficients $c_j$, with $\boldsymbol{c}$ denoting the vector of such coefficients, which we want to determine from experimental measurements. In the context of quantum simulation, the choice of the Ansatz is based on the target Hamiltonian 
one seeks to implement on the quantum device, possibly complemented with additional terms representing deviations from this target Hamiltonian, whose presence we want to test. Reference~\cite{Li2020} proposes to determine the coefficients $c_j$ that yield the best approximation to the implemented Hamiltonian from the condition of energy conservation:
\begin{equation}
  \langle\psi(0)|H|\psi(0)\rangle = \langle\psi(t)|H|\psi(t)\rangle   \quad \forall\,\,\ket{\psi(0)},t\,\,.
\end{equation}
Imposing energy conservation at the level of the Ansatz $H(\boldsymbol{c})$ amounts to choosing $N_{\mathrm{con}}>N_{\mathcal{A}}$ different initial states $\{\ket{\psi_i(0)}\}_{i=1}^{N_{\mathrm{con}}}$ as `constraints', and to minimize the energy differences $|\langle\psi_i(0)|H(\boldsymbol{c})|\psi_i(0)\rangle-\langle\psi_i(t)|H(\boldsymbol{c})|\psi_i(t)\rangle|$ w.r.t.~the coefficients. The optimal $\boldsymbol{c}$, which we denote with $\boldsymbol{c}^{\mathrm{rec}}$ in the following, is thus determined as~\cite{Li2020}
\begin{equation} \label{eq:RecCond}
    \boldsymbol{c}^{\mathrm{rec}} = \mathrm{arg}\,\min_{\boldsymbol{c}} \frac{\Vert M \boldsymbol{c} \Vert}{\Vert \boldsymbol{c} \Vert} \,\,,
\end{equation}
where the elements of the constraint matrix $M$ are given by
\begin{equation}
  \label{eq:Melements}
  M_{i,j} = \langle \psi_i(t) \vert h_j \vert \psi_i(t) \rangle -  \langle \psi_i(0) \vert h_j \vert \psi_i(0) \rangle \,\,,
\end{equation}
which can be inferred from experimental measurements collected from a series of quantum quenches starting from the chosen states $\{\ket{\psi_i(0)}\}_{i=1}^{N_{\mathrm{con}}}$. The \emph{reconstructed Hamiltonian}, denoted with
$H_\mathrm{rec} \equiv H(\boldsymbol{c}^\mathrm{rec})$, is the one that best approximates the implemented $H$ within the Ansatz space spanned by $\mathcal{A}$, in the sense of energy conservation. Comparison of $H_\mathrm{rec}$ with the target Hamiltonian allows one to assess the quality of the quantum device in simulating the model of interest. Note that in Eq.~\eqref{eq:RecCond} the overall scale of $\boldsymbol{c}^\mathrm{rec}$ remains
undetermined, but there exist efficient ways of determining it (see Refs.~\cite{Bairey2019,Evans2019}).

It is essential that the technique above also provides a way of assessing errors, or confidence in the reconstructed parameters $\boldsymbol{c}^\mathrm{rec}$ 
for a given set of experimental data, which are provided here as correlation functions entering the constraint matrix $M$. The quantity that one has access to from HL is the optimal value of the cost function $\Vert M \boldsymbol{c} \Vert$ in Eq.~\eqref{eq:RecCond},
\begin{eqnarray} \label{eq:lambda1}
    \lambda_1 = \frac{\Vert M \boldsymbol{c}^\mathrm{rec} \Vert}{\Vert \boldsymbol{c}^\mathrm{rec} \Vert} = \frac{\sqrt{\sum_i\big|\langle H_{\mathrm{rec}}\rangle_{i,t}-\langle H_{\mathrm{rec}}\rangle_{i,0} \big|^2}}{\Vert\boldsymbol{c}^{\mathrm{rec}}\Vert} \,\,,
\end{eqnarray}
with $\langle H_{\mathrm{rec}}\rangle_{i,t}\equiv\langle \psi_i(t) \vert H_{\mathrm{rec}} \vert \psi_i(t) \rangle$, which measures how well the reconstructed Hamiltonian is conserved during the dynamics governed by $H$. In practice, $\lambda_1$ is calculated as the smallest singular value of $M$, which is shown to be equivalent to the minimum of the `energy cost-function' $\Vert M\boldsymbol{c}\Vert$ in Eq.~\eqref{eq:RecCond} (see Ref.~\cite{Li2020} and \ref{app:HLdetails} for details). 
In Sec.~\ref{subsec:resultsHL} below, $\lambda_1$ will play a key role in our study of the Trotter threshold.

\subsubsection{Learning the Generator of a Trotter Step of the Kicked Top `Order by Order' in the Floquet Magnus Expansion}

We now extend the HL protocol described above to the {\em learning} of the Floquet Hamiltonian of the kicked top, defined via $U_{\tau}= \e^{-\i H_{\mathrm{F}} (\tau) \tau}$ with $U_{\tau}$ given in Eq.~\eqref{eq:U-CUE}. As outlined in the introduction, the kicked top constitutes a particular example of more general DQS of genuinely many-body systems [see Sec.~\ref{subsec:TrotterThreshQMB}], where the HL method proposed here would acquire its most (experimentally) relevant application.

In this context, we choose to rephrase the kicked top as an `experiment' implementing a Trotter evolution cycle which is presented to us as a {\em black box}. We denote the implemented Trotter cycle as $U_{\tau}^{\rm exp}= \e^{-\i H_\mathrm{F}^\mathrm{exp} (\tau) \tau}$, parametrized in terms of the generator $H_\mathrm{F}^\mathrm{exp}(\tau)$. In an ideal experiment $H_\mathrm{F}^\mathrm{exp}(\tau)=H_{\mathrm{F}}(\tau)$, but here we wish to infer the operator content of $H_\mathrm{F}^\mathrm{exp} (\tau)$ from experimental measurements, for two main reasons. First, in DQS we are interested in learning the Trotter errors, contained in $H_\mathrm{F}^\mathrm{exp}(\tau)$ as higher order terms in $\tau$ (see Floquet-Magnus expansion Eq.~\eqref{eq:Fexp}). Second, in an experimental context there might also be control errors, which will be reflected in the operator structure of $H_\mathrm{F}^\mathrm{exp}(\tau)$ deviating from the ideal $H_\mathrm{F}(\tau)$: the reconstruction of $H_\mathrm{F}^\mathrm{exp} (\tau)$ enables the detection of such experimental control errors, and thus a characterization of the Trotter block which we can view as a process tomography of the corresponding quantum circuit \footnote{Our discussion ignores errors due to decoherence, i.e., we assume unitary evolution. The procedure can be extended to the learning of Floquet Liouvillians as it is shown in Ref.~\cite{Pastori2022}.} [see Fig.~\ref{fig:dqs}]. 

To achieve these goals, we seek for a reconstruction of $H_\mathrm{F}^\mathrm{exp} (\tau)$ based on the {\em operator Ansatz}
\begin{equation} \label{eq:Gansatz1}
    H_\mathrm{F}^{\mathrm{exp}\,(K)}(\tau)=\sum_{k=1}^{K} \tau^k \, C^\mathrm{exp}_k \,\,,
\end{equation}
formulated as series in $\tau$ truncated at order $K$, i.e., we learn the generator `order by order' by increasing stepwise $K$. The operator content of the operators $C^\mathrm{exp}_k$ is specified based on our expectation $H_\mathrm{F}(\tau)$ [see the $C_k$ in Eqs.~\eqref{eq:FexpTerms1}-\eqref{eq:FexpTerms3}], but might also contain additional terms whose presence we want to test. Using the notation of the previous subsection, we may rewrite the Ansatz as
\begin{equation} \label{eq:Gansatz2}
    H_\mathrm{F}^{\mathrm{exp}\,(K)}(\tau)=\sum_{h_j\in\mathcal{A}_{K}}c_j\,h_j \,\,,
\end{equation}
with $\mathcal{A}_{K}$ denoting the Ansatz set comprising the operator content of the $C^\mathrm{exp}_k$ up to order $K$ in $\tau$. In this notation, $\mathcal{A}_0$ is the Ansatz for the operator content of the zeroth order terms, corresponding to the generators of the individual gates in $U_{\tau}^{\rm exp}$ (e.g., the $H_{\mu}$ in Eq.~\eqref{eq:U-CUE}), while $\mathcal{A}_{k>0}$ contains all operators generated by the $k$-nested commutators of the terms in $\mathcal{A}_0$, with $k=1,...,K$.

The key quantity representing the quality of our reconstruction is $\lambda_1(\tau)$ introduced in Eq.~\eqref{eq:lambda1}, which quantifies the error of our---typically low order in $\tau$---Ansatz in representing the `experimental' Trotter block. The structure of the Ansatz \eqref{eq:Gansatz2} and the ability of measuring $\lambda_1(\tau)$ for different values of $\tau$ allows us to (i) discriminate between control errors (appearing as `unwanted' terms in $\mathcal{A}_0$) and Trotter errors (the higher order terms), and (ii) detect the value of $\tau_*$ at which these Trotter errors proliferate and quantum chaotic dynamics emerges, corresponding to the stepsize at which our Ansatz \eqref{eq:Gansatz1} fails in approximately capturing the stroboscopic DQS dynamics. This is in analogy to the schematic Fig.~\ref{fig:dqs}(c) where $\lambda_1(\tau)$ will become the proxy for the Hamiltonian distance.


Below, we will apply these ideas to the study of the Trotter threshold in the kicked top, using the behavior of $\lambda_1(\tau)$ as the quantifier signaling the transition from regular to chaotic dynamics when approaching $\tau\rightarrow\tau_{*}$.

\subsection{Trotter Threshold from Hamiltonian Learning: Results for the Kicked Top} \label{subsec:resultsHL}

\begin{figure} 
	\centering  
	\includegraphics[width=\linewidth]{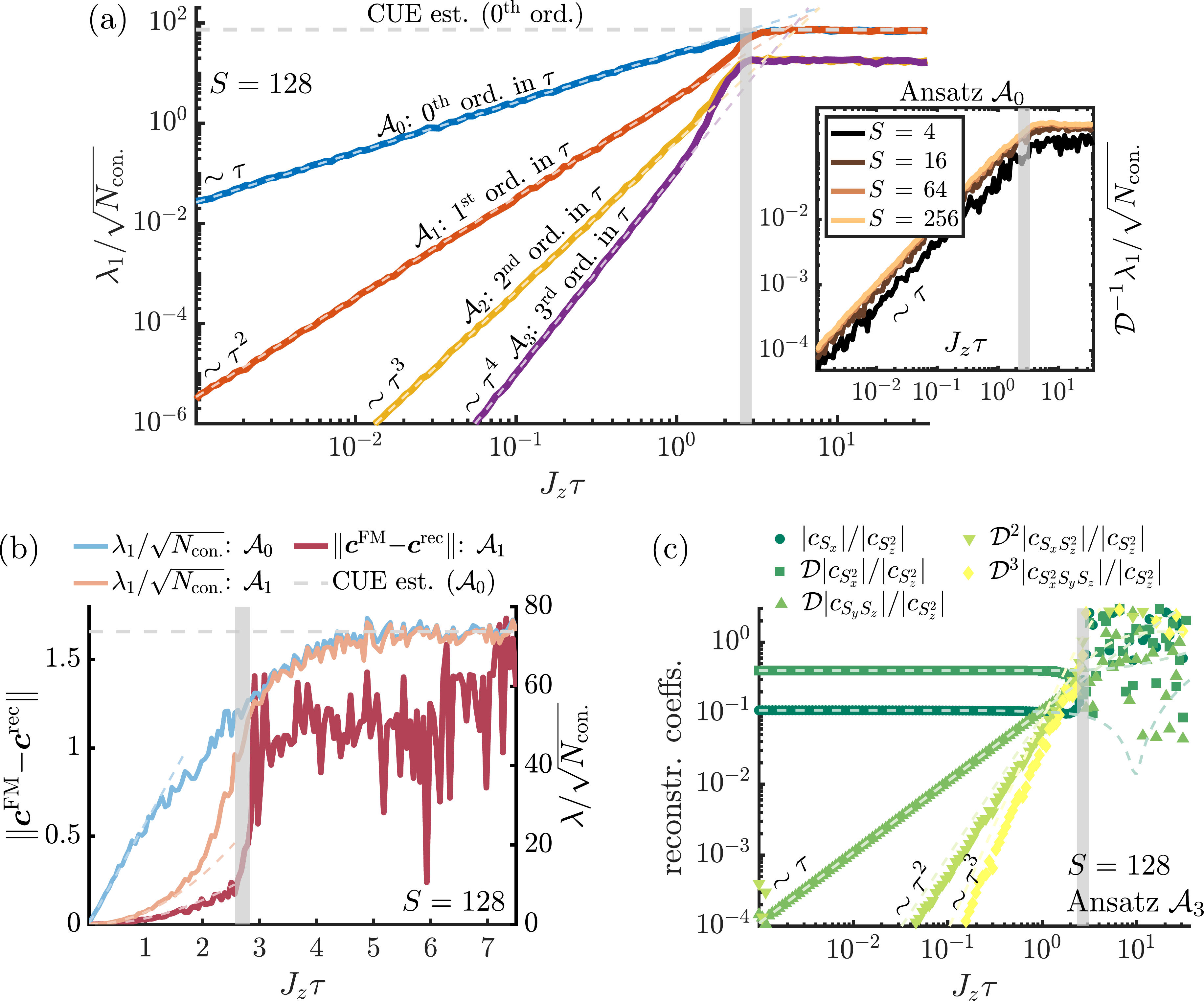}
	\caption{{\bf{Reconstruction of Floquet Hamiltonian of kicked top \eqref{eq:U-CUE}}}. (a) Scaling of $\lambda_1/\sqrt{N_{\mathrm{con}}}$ with $\tau$, for different Ans\"atze corresponding to different truncations (different colors). The Trotter threshold is marked by the vertical shaded line. Inset: dependence of $\lambda_1(\tau)$ on the spin $S$ of the kicked top. (b) Linear plot of the parameter distance $\Vert\boldsymbol{c}^{\mathrm{FM}}-\boldsymbol{c}^{\mathrm{rec}}\Vert$ (dark red solid line), with $\boldsymbol{c}^{\mathrm{FM}}$ and $\boldsymbol{c}^{\mathrm{rec}}$ both normalized to one (in units of $J_z$), compared to $\lambda_1(\tau)$. (c) Some reconstructed parameters (up to $\mathcal{O}(\tau^3)$) compared to the analytical predictions from the FM expansion (dashed lines). The parameters used are $J_x=0.4J_z$, $J_y=0$, $h_x=0.11J_z$, $h_y=h_z=0.1J_z$. In all plots, we chose $\mathcal{D}=2S+1$ random coherent states as initial states for the HL, and evolved the system up to time $100J_z^{-1}$.}
	\label{fig:HLinKT}
\end{figure}

We now present our results for learning the generator $H_\mathrm{F}^\mathrm{exp}(\tau)$ of the Trotter step 
as a function of $\tau$ via HL, as a method to detect and characterize the Trotter threshold in an experimentally feasible protocol. Specifically, we simulate the above protocol for learning the generator of the kicked top dynamics, imagined here as a DQS implementing the Trotter cycle $U_{\tau}$ defined in Eq.~\eqref{eq:U-CUE}. In order to illustrate the main ideas, and to compare to the results of Sec.~\ref{sec:Review}, we consider $U_{\tau}$ without the addition of `unwanted' terms (experimental control errors), i.e., $H_\mathrm{F}^\mathrm{exp}(\tau)=H_{\mathrm{F}}(\tau)$, and simulate the HL protocol in absence of measurement noise in the matrix elements $M_{i,j}$ of Eq.~\eqref{eq:Melements}. To reconstruct $H_{\mathrm{F}}(\tau)$, the Ansatz operators $h_j$ for the HL protocol are chosen such that they capture the operator content of the first few orders of the FM expansion in Eq.~\eqref{eq:Fexp}. We denote with $\mathcal{A}_k$ the set of operators $\{h_j\}_j$ corresponding to a $k$th-order truncation of the FM series. For the kicked top of the present example, the Ansatz sets for the first few orders read as~\footnote{In general, these operators are not hermitian and do not directly correspond to observables, and result in the constraint matrix elements $M_{i,j}$ being complex. However, the matrix elements $M_{i,j}$ can be equivalently constructed as the sum $M_{i,j}=M^{\mathrm{Re}}_{i,j}+\mathrm{i}\,M^{\mathrm{Im}}_{i,j}$, where the expectation values in $M^{\mathrm{Re}}_{i,j}$ and $M^{\mathrm{Im}}_{i,j}$ are taken of hermitian operators $h_j+h_j^{\dagger}$ and $\mathrm{i}(h_j^{\dagger}-h_j)$, respectively.}
\begin{eqnarray*}
& \mathcal{A}_0=\{S_x^2,S_x,S_y,S_z^2,S_z\} \,\,,\\
& \mathcal{A}_1=\mathcal{A}_0\cup\{S_xS_y, S_yS_z, S_xS_z, S_xS_yS_z\} \,\,,\\
& \mathcal{A}_2=\mathcal{A}_1\cup\{S_x^2S_y, S_yS_z^2, S_x^2S_z^2, S_z^4, S_x^2S_z, S_z^3, S_xS_z^2, S_x^4, S_x^3\} \,\,,
\end{eqnarray*}
containing all the linearly independent products of spin operators coming from the commutators in Eq.~\eqref{eq:FexpTerms}~\footnote{We typically choose $k\ll S$, hence the operators $h_j$ are low powers of collective spin operators, which constitute few-body operators when interpreting the kicked top as an ensemble of spin-$1/2$ degrees of freedom.}. The elements $M_{i,j}$ of the constraint matrix $M$ defined in Eq.~\eqref{eq:Melements} are determined from expectation values of these operators over the states $\ket{\psi_i(n\tau)}=U_{\tau}^n\ket{\psi_i(0)}$. To compare to the results of Sec.~\ref{sec:Review}, we choose $n\tau=100J_z^{-1}$ and random coherent states as initial states: our results are however independent of these specific choices. The Trotter threshold discussed in Sec.~\ref{sec:Review} will be revealed by measuring $\lambda_1$ for several values of the Trotter step $\tau$, keeping the Ansatz $\mathcal{A}_k$ fixed as $\tau$ is changed. Specifically, the behavior of $\lambda_1(\tau)$ serves as an indicator of the quality of HL, which is bound to fail for $\tau>\tau_{*}$ where the dynamics of the kicked top cannot be approximated by a truncated FM expansion.

The behavior of $\lambda_1$ as a function of $\tau$ is exemplified in Fig.~\ref{fig:HLinKT}(a) for several $\mathcal{A}_k$. Two different behaviors are clearly visible, separated by a grey vertical line denoting the estimated value of $\tau_{*}$, corresponding to the Trotter threshold discussed in Sec.~\ref{sec:Review}. Let us now describe these regimes and relate them to the observations presented in Sec.~\ref{sec:Review}.

\subsubsection{Pre-Threshold Regime.} 
For $\tau<\tau_{*}$ we see that, depending on the Ansatz, $\lambda_1$ obeys different scaling with $\tau$, i.e., $\lambda_1\in\mathcal{O}(\tau^{k+1})$ for a $k$th-order Ansatz $\mathcal{A}_k$. This behavior of $\lambda_1(\tau)$ confirms that for $\tau<\tau_{*}$ a low-order truncation of the FM expansion is sufficient to accurately describe the stroboscopic dynamics of the kicked top. Indeed, it is intuitively clear that an Ansatz $\mathcal{A}_k$ capturing only the lowest $k$ orders of $H_{\mathrm{F}}(\tau)$ results in a reconstructed Hamiltonian $H_{\mathrm{rec}}(\tau)=\sum_jc^{\mathrm{rec}}_j(\tau)h_j$ violating energy conservation by terms of order $\mathcal{O}(\tau^{k+1})$ during the stroboscopic dynamics, i.e, $\Vert H_\mathrm{rec}(\tau) - H_{\mathrm{F}}(\tau) \Vert \in \mathcal{O}(\tau^{k+1})$. Since $\lambda_1$ is precisely what captures how well $H_{\mathrm{rec}}(\tau)$ is conserved [see Eq.~\eqref{eq:lambda1}], the scaling of $\Vert H_\mathrm{rec}(\tau) - H_{\mathrm{F}}(\tau) \Vert$ must be reflected in that of $\lambda_1(\tau)$, hence $\lambda_1\in\mathcal{O}(\tau^{k+1})$. This is also confirmed by the behavior of the \emph{parameter distance} $\Vert\boldsymbol{c}^{\mathrm{FM}}-\boldsymbol{c}^{\mathrm{rec}}\Vert$, shown by the dark red line in Fig.~\ref{fig:HLinKT}(b), where $\boldsymbol{c}^{\mathrm{FM}}$ denotes the vector of coefficients that are calculated from the analytical FM expansion (truncated to order $k$ for an Ansatz $\mathcal{A}_k$). As long as $\tau<\tau_{*}$, $\Vert\boldsymbol{c}^{\mathrm{FM}}-\boldsymbol{c}^{\mathrm{rec}}\Vert$ remains small and its behavior as a function of $\tau$ reflects that of $\lambda_1(\tau)$.

\subsubsection{Trotter Threshold}\label{subsubsec:TrotterThreshold}
At $\tau\approx\tau_{*}$, $\lambda_1(\tau)$ transitions from a $\mathcal{O}(\tau^k)$ behavior to a $\tau$-independent value. The reason of this transition is the fact that a low-order truncation of $H_{\mathrm{F}}(\tau)$ becomes insufficient to accurately describe the dynamics. More precisely, since our HL protocol is based on monitoring the dynamics of few-body observables (the $h_j\in\mathcal{A}_k$) which are measured in order to construct the matrix $M$, the onset of the plateau in $\lambda_1(\tau)$ signals the proliferation of Trotter errors in the dynamics of such observables. The behavior of $\lambda_1(\tau)$ is therefore analogous to that of the simulation accuracy $\overline{Q_E}(\tau)$, defined in Eq.~\eqref{eq:QE}, showing the threshold behavior at the same value $\tau_{*}$. In fact, $\overline{Q_E}(\tau)$ and $\lambda_1(\tau)$ encode very similar physical information, in that they both are related to how well energy, as defined by the time-averaged Hamiltonian in one case and by a low-order truncation of FM expansion in the other, is conserved. Specifically, it is clear from its definition [Eq.~\eqref{eq:lambda1}] that $\lambda_1(\tau)$
can be seen as a simulation accuracy $Q_E(n\tau)$ [Eq.~\eqref{eq:QE}] averaged over several initial states, upon identifying the energy $E_{\tau}$ with the reconstructed Hamiltonian $H_{\mathrm{rec}}(\tau)$ (up to an overall normalization). 
We notice that threshold behavior can be observed also in the parameter distance defined before, and shown in Fig.~\ref{fig:HLinKT}(b), as well as in the reconstructed Hamiltonian parameters which, while showing almost perfect agreement with the analytically determined FM expansion for $\tau<\tau_{*}$, become essentially random for $\tau>\tau_{*}$. We show the behavior with $\tau$ of some of the reconstructed parameters in Fig.~\ref{fig:HLinKT}(c). Furthermore, we observe that the value of $\tau_{*}$ does not strongly depend on the spin size $S$, as we show in the inset of Fig.~\ref{fig:HLinKT}(a). This is consistent with the results of our previous work \cite{Sieberer2019}, where we observed that $\tau_{*}$ remained finite even in the classical limit $S\to\infty$.

\subsubsection{Post-Threshold Regime}
For $\tau>\tau_{*}$ we observe a constant value of $\lambda_1(\tau)$, independently of the Ansatz chosen. As shown in Sec.~\ref{sec:Review}, the regime $\tau>\tau_{*}$ corresponds to the chaotic regime where the spectral statistics of $U_{\tau}$ is faithful to RMT. As a consequence, the value of $\lambda_1(\tau>\tau_{*})$ shows perfect agreement with the value of $\lambda_1$ that can be estimated using RMT. In this sense, $\lambda_1(\tau)$ can be seen as an indicator of the onset of chaos, showing a behavior similar to the level spacing ratio $r$ and the PR introduced in Sec.~\ref{sec:Review}. Here, the RMT estimate of $\lambda_1$ has to be understood as the value that one would obtain when carrying out the HL protocol above with the same Ansatz $\mathcal{A}_k$, but replacing $U_{\tau}$ with a random $U$ drawn from the relevant matrix ensemble (CUE for the example presented here). We refer the reader to \ref{app:RMTlambda} for the details of this calculation. In the plots, only the CUE estimate for the zeroth order Ansatz is shown, corresponding to the horizontal grey dashed line in panels (a) and (b): changing the Ansatz to reconstruct higher orders of the FM expansion leads to different values for the CUE estimate, which match with the observed plateaus in the plots.

\section{Conclusions and Outlook} \label{sec:Outlook}
In the present paper, we have shown how Hamiltonian learning can be employed as a method for analyzing the transition to chaotic dynamics in the kicked top, interpreted here as a DQS of a collective spin system where this transition manifests itself in the proliferation of Trotter errors for a certain Trotter step size $\tau_{*}$ \cite{Sieberer2019}. In general, our method gives a recipe for interrogating (via measurements) a quantum device implementing a Trotterized evolution, asking the question: can the implemented dynamics be approximately described by a time-independent \emph{few-body} Hamiltonian? A low-order truncation of the FM expansion constitutes an example of such a Hamiltonian, whose structure motivates our choice of the Ansatz for HL, and thus of the measurements performed to interrogate the system. The answer to the question above is provided by $\lambda_1$, assessing the quality of the Ansatz: by measuring $\lambda_1$ as a function of $\tau$ one can detect the value $\tau_{*}$ at which a description in terms of a few-body Hamiltonian breaks down, corresponding to the proliferation of Trotter errors in DQS.

When applied to the kicked top, our protocol requires only the measurement of products of few collective spin operators, and may be readily implemented in experimental setups where the kicked top dynamics can be realized. These include existing realizations of the kicked top as the spin of single atoms \cite{Chaudhury2009}, nuclear spins \cite{Mourik2018}, composite spins in
nuclear magnetic resonance \cite{Krithika2019}, and collective spins of ensembles of superconducting qubits \cite{Neill2016}, together with potential realizations with magnetic atoms \cite{Baier2018,Chalopin2018} and trapped ions \cite{Britton2012,Garttner2017}.

We emphasize that the method presented here applies also to the case of DQS of genuinely many-body systems \cite{Pastori2022}, of which the kicked top can be seen as a limit where the dynamics, for suitable initial conditions, admits a description in terms of collective spin operators [see Sec.~\ref{subsec:TrotterThreshQMB}]. In view of the experimental applicability of our HL protocol in this more general context, we conclude by commenting on its relation to the previously proposed signatures of Trotter threshold and chaos transition in Trotterized dynamics.

First, while we have pointed out a similarity between the physical information encoded in $\lambda_1$ and the simulation accuracy $Q_E$ [see Sec.~\ref{subsubsec:TrotterThreshold}], we note that there exist a fundamental difference between the two. The definition of the simulation accuracy is reliant on the exact knowledge of the implemented Hamiltonian which, in an experimental setting, may be unknown or only approximately known. Conversely, the value of $\lambda_1$ is determined \emph{from the reconstruction of the implemented Hamiltonian}, and hence does not rely on previous knowledge of Hamiltonian parameters. Thus, measuring $\lambda_1(\tau)$ simultaneously achieves the goal of (i) characterizing a DQS in terms of the implemented stroboscopic Hamiltonian, and (ii) detecting the regime in which Trotter errors on few-body observables are controlled in $\tau$.

Second, we point out that while the participation ratio and level spacing ratio studied in Sec.~\ref{sec:Review} are standard ways of probing the notion of quantum chaos, their measurement in the context of general quantum many-body systems is non-trivial. In this context, we would like to mention recent advances in this direction \cite{Joshi2021} based on the use of randomized measurements \cite{Elben2017,Vermersch2018,Vermersch2018a}. Conversely, although the behavior of $\lambda_1$ is only an indicator of the onset of chaos, it is measurable with costs that in general scale polynomially in the system size. 

Finally, we mention that even if the results presented here do not contain any simulated experimental imperfections nor measurement shot noise, the presented protocol retains its validity as long as the dynamics of the system can be treated as approximately unitary for the relevant timescales. Furthermore, via the value of $\lambda_1$, our method provides means of assessing the quality of this approximation. A detailed analysis of the influence of measurement noise and imperfections is beyond the scope of this work, and we refer the reader to Ref.~\cite{Pastori2022} for a detailed study of realistic applications of this protocol in many-body DQS.

While the emphasis of the present work has been on connecting the framework of Hamiltonian learning with the regular-chaotic dynamics of the kicked top in the more general context of DQS of many-body systems, we believe that these concepts and techniques can also be used to shed new light on quantum chaos in the kicked top per se, a field that has seen Fritz Haake as one of its most active and brilliant contributors.

\section*{Acknowledgments}

We thank Lata K.~Joshi, Markus Heyl, Philipp Hauke, Nathan Langford and Cahit Kargi for valuable discussions. This work was supported by Simons Collaboration on Ultra-Quantum
Matter, which is a grant from the Simons Foundation
(651440, P.Z.), the European Union’s Horizon 2020 research and innovation programme under Grant Agreement No.~817482 (Pasquans), and by LASCEM by
AFOSR No.~64896-PH-QC. L.M.S.\ acknowledges support from the Austrian Science Fund (FWF): P 33741-N. The computational results presented here have been obtained (in part) using the LEO HPC infrastructure of the University of Innsbruck.

\section*{Data Availability}
The data sets generated and analyzed during the current study are available from the
corresponding author on reasonable request.

\appendix

\section{Variants  of the kicked top model}
\label{app:vari-kick-top}

The time-dependent Hamiltonian in Eq.~\eqref{eq:H-kicked-top} describes the
prototypical and---with $H_x$ being purely linear and $H_z$ purely quadratic in
spin operators---simplest version of the kicked top capable of global
chaos~\cite{Haake2018}. Simplicity brings about a high degree of symmetry. This
is best discussed in terms of the Floquet operator Eq.~\eqref{eq:U-kicked-top}:
Time-reversal covariance of $U_{\tau}$, which is expressed through the relation
$T U_{\tau} T^{\dagger} = U_{\tau}^{\dagger}$ where
$T = \e^{\i H_x \tau} K$ and $K$ is complex conjugation, implies
that $U_{\tau}$ belongs to the COE. Further, $U_{\tau}$ is symmetric under
rotations by $\pi$ around the $x$ axis, $R_x U_{\tau} R_x^{\dagger} = U_{\tau}$
with $R_x = \e^{\i \pi S_x}$. Hence, an RMT analysis, e.g., of level spacings,
has to be carried out independently within the subspaces of eigenstates of
$U_{\tau}$ that are even and odd under $R_x$, respectively.  Finally, there are
isolated points of even higher symmetry: if $h_x \tau$ is a multiple of $2 \pi$,
then the entire spectrum of $H_x \tau = h_x \tau S_x$ contains only multiplies
of $2 \pi$, such that $\e^{- \i H_x \tau} = 1$ and the Floquet operator reduces
to $U_{\tau} = \e^{- \i H_z \tau}$. Consequently, $S_z$ is conserved, and the
dynamics becomes trivially integrable. These points of resonant driving are
necessarily encountered when the driving period $\tau$ is varied over a wide
range, as we do when we study the Trotter threshold, and, while being well
understood, they somewhat spoil the phenomenology of the threshold
behavior. Further, in view of the fact that the Trotter threshold occurs
in a variety of models~\cite{Langford2016}, the existence of such resonant
driving points may be regarded as an artefact of the choice of $H_x$ having an
equidistant spectrum. Therefore, in Ref.~\cite{Sieberer2019} and in the main
text, we focus on the more generic situation, in which $H_x$ is augmented by a
term that is quadratic in $S_x$. Further, by adding a term that is linear in
$S_z$ to $H_z$, we rid the Floquet operator of geometric symmetries, and thus
simplify the RMT analysis. Finally, the RMT analysis of HL presented in
\ref{app:RMTlambda}, while proceeding along the same lines, simplifies
considerably for the CUE as compared to the COE. Hence, to not overburden the
presentation of our results with unnecessary technicalities, we choose to focus
on a version of the kicked top without time reversal symmetry, which is given in
Eq.~\eqref{eq:U-CUE}~\cite{Haake2018}.

\section{Details on the Hamiltonian Learning Protocol of Sec.~\ref{sec:HLinKickedTop}}
\label{app:HLdetails}
In this appendix, we provide additional details on the HL protocol explained in Sec.~\ref{subsec:HL}, showing how the reconstruction of the Hamiltonian parameters is achieved via singular value decomposition (SVD).
We write the Ansatz $H(\boldsymbol{c})$ for the Hamiltonian to be reconstructed as $H(\boldsymbol{c})=\sum_{j=1}^{N_{\mathcal{A}}} c_j\,h_j$, specified by a chosen set $\mathcal{A}$ of Ansatz operators $\mathcal{A}=\{h_j\}_{j=1}^{N_{\mathcal{A}}}$, and we want to find the optimal coefficients $\boldsymbol{c}^{\mathrm{rec}}$ by solving
\begin{equation}
    \boldsymbol{c}^{\mathrm{rec}}=\mathrm{arg}\,\min_{\boldsymbol{c},\Vert\boldsymbol{c}\Vert=1}\sum_{i=1}^{N_{\rm con}}\big|\langle H(\boldsymbol{c})\rangle_{i,t}-\langle H(\boldsymbol{c})\rangle_{i,0}\big|^2 \,\,, 
\end{equation}
with $\langle H(\boldsymbol{c})\rangle_{i,t}=\langle \psi_i(t)|H(\boldsymbol{c})|\psi_i(t)\rangle$, for a set of $N_{\rm con}$ chosen initial states $|\psi_i(0)\rangle$. This is equivalent to
\begin{equation} \label{aeq:minimize_normMc}
    \boldsymbol{c}^{\mathrm{rec}}=\mathrm{arg}\,\min_{\boldsymbol{c},\Vert\boldsymbol{c}\Vert=1}\Vert M\boldsymbol{c}\Vert^2 \,\,, 
\end{equation}
with the constraint matrix $M$ having elements defined in Eq.~\eqref{eq:Melements}, and amounts to determining $\boldsymbol{c}^{\mathrm{rec}}$ as the right singular vector of $M$ corresponding to the smallest singular value $\lambda_1$, that is 
\begin{equation} \label{aeq:RecEq}
    M\boldsymbol{c}^\mathrm{rec} = \lambda_1\boldsymbol{w}_1 \,\,,
\end{equation}
where $\boldsymbol{w}_1$ is the corresponding left singular vector with $\Vert\boldsymbol{w}_1\Vert=\Vert \boldsymbol{c}^\mathrm{rec}\Vert$. This can be seen as follows: from the SVD of $M$ one writes $M=WSV^{\dagger}$, with $W$ having $N_{\mathcal{A}}$ orthonormal columns $\boldsymbol{w}_n$ (each being an $N_{\mathrm{con}}$-component vector), $V^{\dagger}$ having $N_{\mathcal{A}}$ orthonormal rows $\boldsymbol{v}_n^{\dagger}$ (each being an $N_{\mathcal{A}}$-component vector), and $S=\mathrm{diag(\lambda_1,...,\lambda_{N_{\mathcal{A}}})}$, with $0\le\lambda_1\le...\le\lambda_{N_{\mathcal{A}}}$. Thus $M\boldsymbol{c}=WSV^{\dagger}\boldsymbol{c}=\sum_n\lambda_n\,(\boldsymbol{v}_n\cdot\boldsymbol{c})\,\boldsymbol{w}_n$, with $\cdot$ denoting the scalar product, which yields
\begin{equation}
\Vert M\boldsymbol{c}\Vert^2=\sum_n\lambda_n^2\,(\boldsymbol{v}_n\cdot\boldsymbol{c})^2\ge\lambda_1^2(\boldsymbol{v}_1\cdot\boldsymbol{v}_1)^2=\lambda_1^2 \,\,,
\end{equation}
where the absolute minimum is obtained for $\boldsymbol{c}=\boldsymbol{v}_1$. Thus, $\boldsymbol{c}^{\mathrm{rec}}=\boldsymbol{v}_1$.

\section{Random matrix theory estimate for $\lambda_1$}
\label{app:RMTlambda}

In this appendix, we show that for $U_{\tau}=U\in\mathrm{CUE}$ one can make analytical predictions for the form of the matrix $Q=\frac{1}{N_{\mathrm{con}}}M^{\top}M$, thereby efficiently estimating the value of the $\lambda_1$. The result is in good agreement with the value obtained from our HL protocol for large $\tau$ and large number of driving cycles. We start by calculating the expectation value of the elements $Q_{j,k}=\frac{1}{N_{\mathrm{con}}}\sum_iM_{i,j}M_{i,k}$ over the CUE as
\begin{equation} \label{aeq:Q_expRME}
    Q^{\mathrm{CUE}}_{j,k}=\frac{1}{N_{\mathrm{con}}}\sum_i\mathbbm{E}_{U\in\mathrm{CUE}}\big[M_{i,j}M_{i,k}\big]\equiv\frac{1}{N_{\mathrm{con}}}\sum_iq^{\mathrm{CUE}}_{j,k}[\rho_i]\,\,,
\end{equation}
with $\rho_i$ denoting the chosen initial states for the protocol (which in general can be also mixed). The result depends on the chosen set $\{\rho_i\}$, and can be calculated analytically in the case considered in the main text, namely the $\rho_i$ being coherent states. Expressing the matrix elements $M_{i,j}$ as $M_{i,j}=\mathrm{tr}(\rho_i h_j)-\mathrm{tr}(\rho_iU^{\dagger}h_jU)\big)$, and using the methods developed in \cite{Brower1996}, we obtain
\begin{eqnarray} 
    q^{\mathrm{CUE}}_{j,k}[\rho_i]=&\mathrm{tr}(\rho_i h_j)\mathrm{tr}(\rho_i h_k)-\frac{\mathrm{tr}(h_j)}{\mathcal{D}}\mathrm{tr}(\rho_i h_k) \nonumber \\
    &-\frac{\mathrm{tr}(h_k)}{\mathcal{D}}\mathrm{tr}(\rho_i h_j)
    +\frac{\mathrm{tr}(h_j)\mathrm{tr}(h_k)}{\mathcal{D}^2-1}+\frac{\mathrm{tr}(\rho_i^2)\mathrm{tr}(h_jh_k)}{\mathcal{D}^2-1} \nonumber \\
    &-\frac{\mathrm{tr}(h_jh_k)}{\mathcal{D}(\mathcal{D}^2-1)}-\frac{\mathrm{tr}(\rho_i^2)\mathrm{tr}(h_j)\mathrm{tr}(h_k)}{\mathcal{D}(\mathcal{D}^2-1)} \,\,,
    \label{aeq:QelemCUE}
\end{eqnarray}
where $\mathcal{D}$ is the Hilbert space dimension, i.e., $\mathcal{D}=2S+1$. Since we use pure initial states, $\mathrm{tr}(\rho_i^2)=1$. In the case of the kicked top, the Ansatz operators $h_j$ are products of spin operators, and we can use the results of Ref.~\cite{Ambler1962} to obtain the analytical expressions for the traces $\mathrm{tr}(h_j)$ and $\mathrm{tr}(h_jh_k)$. 
where $\mu,\nu=x,y,z$. The expectation values $\mathrm{tr}(\rho_ih_j)$ are easily calculated for the chosen initial states. If we are interested in the zeroth order terms in the Floquet Hamiltonian \eqref{eq:Fexp}, we can use the following identities: $\langle\theta,\phi|\boldsymbol{S}|\theta,\phi\rangle=S(\sin\theta\cos\phi,\sin\theta\sin\phi,\cos\theta)$, $\langle\theta,\phi|S_{x}^2|\theta,\phi\rangle=S\big(S-\frac{1}{2}\big)(\sin\theta\cos\phi)^2+\frac{S}{2}$, and $\langle\theta,\phi|S_{z}^2|\theta,\phi\rangle=\frac{S}{4}(1+2S+(2S-1)\cos 2\theta)$ with $\boldsymbol{S}=(S_x,S_y,S_z)$. For a large number of initial states, the sum $N_{\mathrm{con}}^{-1}\sum_i$ can be approximated by an expectation value over the initial states ensemble (ISE), i.e., 
\begin{equation} \label{aeq:ISEexp_Q}
    \frac{1}{N_{\mathrm{con}}}\sum_iq^{\mathrm{CUE}}_{j,k}[\rho_i]\approx\mathbbm{E}_{\rho\in\mathrm{ISE}}\,q^{\mathrm{CUE}}_{j,k}[\rho]\equiv\mathcal{Q}_{j,k} \,\,.
\end{equation}
which, for coherent states (ISE=$\mathrm{CS}$), is calculated using $\mathbbm{E}_{\rho\in\mathrm{CS}}\big[\mathrm{tr}(\rho h_j)\big]=\frac{1}{4\pi}\int\mathrm{d}\Omega_{\theta,\phi}\langle h_j\rangle_{\theta,\phi}$ and $\mathbbm{E}_{\rho\in\mathrm{CS}}\big[\mathrm{tr}(\rho h_j)\mathrm{tr}(\rho h_k)\big]=\frac{1}{4\pi}\int\mathrm{d}\Omega_{\theta,\phi}\langle h_j\rangle_{\theta,\phi}\langle h_k\rangle_{\theta,\phi}$
with $\langle h_j\rangle_{\theta,\phi}=\langle \theta,\phi|h_j|\theta,\phi\rangle$ and $\int\mathrm{d}\Omega_{\theta,\phi}=\int_0^{2\pi}\mathrm{d}\phi\int_0^{\pi}\mathrm{d}\theta\sin\theta$. Using Eqs.~\eqref{aeq:Q_expRME}, \eqref{aeq:QelemCUE} and \eqref{aeq:ISEexp_Q}, we can obtain an analytical expression for the matrix $\mathcal{Q}$ which, in the case of Ansatz set $\mathcal{A}_0=\{S_z^2,S_z,S_y,S_x^2,S_x\}$, reads as (we show only the non-zero matrix elements)
\begin{eqnarray*}
    &\mathcal{Q}_{S_z^2,S_z^2}=\mathcal{Q}_{S_x^2,S_x^2}=\frac{S(8S^3-4S^2+6S-3)}{90}\,\,,\\
    &\mathcal{Q}_{S_z^2,S_x^2}=-\frac{S(8S^3-4S^2+6S-3)}{180}\,\,,\\
    &\mathcal{Q}_{S_x,S_x}=\mathcal{Q}_{S_y,S_y}=\mathcal{Q}_{S_z,S_z}=\frac{S(2S+1)}{6}\,\,,
\end{eqnarray*}
from which we estimate the RMT value of $\lambda_1$, denoted with $\lambda_{\mathrm{RMT}}$, as
\begin{equation}
    \lambda_{\mathrm{RMT}}=\sqrt{\varepsilon_1(\mathcal{Q})} \,\,,
\end{equation}
where $\varepsilon_1(\mathcal{Q})$ denotes the lowest eigenvalue of $\mathcal{Q}$. This yields the estimates used in the main text.

As a final remark we emphasize that, while here we are estimating $\lambda_1$ as $\lambda_{\mathrm{RMT}}=\sqrt{\varepsilon_1(\mathbbm{E}_{\mathrm{ISE},\mathrm{CUE}}\,Q)}$, the correct prediction would be $\lambda_{\mathrm{RMT}}=\mathbbm{E}_{\mathrm{ISE},\mathrm{CUE}}\sqrt{\varepsilon_1(Q)}$. However (i) it is difficult to make analytical progress in the second case and (ii) the two calculations agree very well for large Hilbert space dimension $\mathcal{D}$ and large number $N_{\mathrm{ini}}$ of initial states used. Indeed, the variance of the $Q_{j,k}$ over the CUE, $\mathrm{Var}_{\mathrm{CUE}}Q_{j,k}\equiv N_{\mathrm{con}}^{-1}\sum_i\mathrm{Var}_{\mathrm{CUE}}\big[M_{i,j}M_{i,k}\big]$, which can also be analytically calculated using the methods in \cite{Brower1996}, is suppressed in $\mathcal{D}$ as $\mathcal{O}(\mathcal{D}^{-3})$. Furthermore, the variance of $Q_{j,k}$ over the ISE, estimated as $\mathrm{Var}_{\mathrm{ISE}}Q_{j,k}\equiv\mathrm{Var}_{\rho\in\mathrm{ISE}}\,q^{\mathrm{CUE}}_{j,k}[\rho]$, results in random fluctuations of each realization of $Q$ around the mean value $\mathcal{Q}$, which are suppressed as $N_{\mathrm{ini}}^{-1/2}$. That is to say, each realization of $Q$ for a given random unitary and a given set of $N_{\mathrm{con}}$ random initial states can be written as $Q=\mathcal{Q}+R_{\mathrm{V}}$ where $R_{\mathrm{V}}$ is a random matrix whose elements can be approximated as
\begin{equation*}
    (R_{\mathrm{V}})_{j,k}\approx (R_{\mathrm{V}}(\xi,\zeta))_{j,k}=\xi\sqrt{\frac{\mathrm{Var}_{\mathrm{ISE}}Q_{j,k}}{N_{\mathrm{con}}}}+\zeta\sqrt{\mathrm{Var}_{\mathrm{CUE}}Q_{j,k}} \,\,,
\end{equation*} 
with $\xi,\zeta\sim\mathcal{N}(0,1)$, if we treat the corrections coming from the statistical fluctuations over the CUE and ISE as independent and normal distributed random variables. Thus, $R_{\mathrm{V}}$ can be seen as a small perturbation to $Q$ which, in general, has only perturbative effects on its eigenvalues. In principle, one could achieve a more accurate estimate of $\lambda_1$ by calculating $\lambda_{\mathrm{RMT}}=\mathbbm{E}_{\xi,\zeta\sim\mathcal{N}(0,1)}\,\sqrt{\varepsilon_1\big[\mathcal{Q}+R_{\mathrm{V}}(\xi,\zeta)\big]}$, which can also be done efficiently. \\

\vspace{1cm}

\bibliography{HaakeMemorial.bib}

\end{document}